\documentclass[aps,prl,twocolumn,superscriptaddress]{revtex4-2}
\usepackage{graphicx}
\usepackage{latexsym}
\usepackage{amssymb}
\usepackage{amsmath}
\usepackage{amsfonts}
\usepackage[vcentermath]{youngtab}
\usepackage{upgreek}
\usepackage{bm,dsfont}
\usepackage{multirow}
\usepackage{enumitem}
\usepackage{color}
\usepackage{hyperref}
\hypersetup{
colorlinks = true,
linkcolor = [rgb]{0.70,0.13,0.13},
citecolor = [rgb]{0.13,0.55,0.13},
urlcolor = [rgb]{0.25, 0.41, 0.88}}
\usepackage{makecell}

\newcommand{\ket}[1]{|#1 \rangle}

\newcommand{\dd}{\mathrm{d}}

\newcommand{\ii}{\mathrm{i}}
\newcommand{\e}{\mathrm{e}}

\newcommand{\U}{\mathrm{U}}

\newcommand{\dsZ}{\mathbb{Z}}

\newcommand{\scD}{\mathcal{D}}

\newcommand{\scK}{\mathcal{K}}

\newcommand{\scS}{\mathcal{S}}

\newcommand{\Tr}{\operatorname{Tr}}

\renewcommand{\Re}{\operatorname{Re}}
\renewcommand{\Im}{\operatorname{Im}}
\newcommand{\vect}[1]{{\bm{#1}}}

\newcommand{\eq}[1]{\begin{equation}#1\end{equation}}
\newcommand{\eqs}[1]{\begin{equation}\begin{split}#1\end{split}\end{equation}}
\newcommand{\eqnref}[1]{Eq.\,\eqref{#1}}
\newcommand{\figref}[1]{Fig.\,\ref{#1}}

\newcommand{\refcite}[1]{Ref.\,\onlinecite{#1}}

\begin{document}

\title{Optical Conductivity in Symmetric Mass Generation Insulators}

\author{Meng Zeng}
\affiliation{Department of Physics, University of California at San Diego, La Jolla, CA 92093, USA}
\author{Fu Xu}
\affiliation{Department of Physics, Nanjing University, Nanjing, Jiangsu 210093, China}
\author{Da-Chuan Lu}
\affiliation{Department of Physics, University of California at San Diego, La Jolla, CA 92093, USA}
\author{Yi-Zhuang You}
\email{yzyou@physics.ucsd.edu}
\affiliation{Department of Physics, University of California at San Diego, La Jolla, CA 92093, USA}
\date{\today}

\begin{abstract}
Symmetric mass generation (SMG) insulators are interaction-driven, featureless Mott insulating states in quantum many-body fermionic systems. Recent advancements suggest that zeros in the fermion Green's function could lead to non-vanishing negative optical conductivity in SMG insulators, even below the charge excitation gap. This study explores the origin of this unusual behavior through the lens of pole-zero duality, highlighting a critical issue where the current operator becomes unbounded, rendering the response function unphysical. By employing a lattice model, we derive a well-behaved lattice regularization of the current operator, enabling a detailed study of optical conductivity in SMG insulators. Utilizing both analytical and numerical methods, including strong-coupling expansions, we confirm that SMG insulators exhibit no optical conductivity at low energies below the charge gap, effectively resolving the paradox. This work not only deepens our understanding of quantum many-body phenomena but also lays a robust theoretical groundwork for future experimental explorations of SMG materials.
\end{abstract}

\maketitle

\emph{Introduction.}~---
Symmetric mass generation (SMG) insulators \cite{Wang1307.7480,Ayyar1410.6474,Slagle1409.7401,BenTov1412.0154,Catterall1510.04153,Ayyar1511.09071,Ayyar1606.06312,Witten1605.02391,Ayyar1611.00280,He1603.08376,DeMarco1706.04648,Ayyar1709.06048,You1711.00863,Schaich1710.08137,Kikukawa1710.11618,Kikukawa1710.11101,Butt1811.01015,Butt1810.06117,Wang1809.11171,Catterall2002.00034,Razamat2009.05037,Catterall2010.02290,Butt2111.01001,Zeng2202.12355,Hou2212.13364,Lu2210.16304,Lu2302.12731,Guo2023S2306.17420,Lu2023S2308.11195,Liu2024D2308.07380} represent a novel class of interaction-driven \emph{featureless} Mott insulating states in quantum many-body systems of fermions. These systems feature the cancellation of all quantum anomalies \cite{Tong2104.03997,Butt2101.01026,Wang2204.14271}, such that a symmetric gapped state without spontaneous symmetry breaking (SSB) or topological order is allowed, which we refer to as ``featureless''. SMG insulators are characterized by a full energy gap to all fermionic and bosonic excitations, including collective charge and current excitations. The excitation gap arises from non-perturbative interaction effects and eludes mean-field theoretical explanations.

Central to the theoretical understanding of SMG insulators is the behavior of the fermion Green’s function, defined as:
$G(k):=-\langle\psi_{k}\psi_{k}^\dagger\rangle$,
where \(\psi_k\) is the fermion operator at the energy-momentum \(k=(\omega,\vect{k})\). A key aspect of the SMG insulator is that its fermion Green's function determinant approaches zero as energy-momentum tends to zero \cite{Gurarie1011.2273,You1403.4938,Catterall1609.08541,You1705.09313,Catterall1708.06715,Xu2103.15865,Lessnich2021E2103.02624,Hu2021T2110.06182,Kaplan2021I2112.06954,Kaplan2022G2205.05707,Setty2023S2301.13870,Chang2023F2311.09970,Lu2023G2307.12223,Setty2023T2311.12031,Chen2024D2401.12156,Golterman2024P2311.12790}, i.e.
\eq{
\det G(k)=0\text{ as } k_\mu k^\mu\to 0.}
This property, the zero of the Green’s function, raises an intriguing question regarding its experimental implications \cite{Lu2023G2307.12223,Lu2023S2308.11195,Zhang2023S2309.05726,Chang2023F2311.09970,Setty2023T2311.12031,Chen2024D2401.12156}. Recently, Golterman and Shamir \cite{Golterman2024P2311.12790} proposed that these zeros might significantly influence the electromagnetic response of the SMG insulator when the fermions are coupled to a background $\U(1)$ gauge field, particularly suggesting a scenario where the SMG insulator exhibits non-vanishing charge conductivity at low energy, even below the energy gap of all excitations.

This apparent puzzle of how an \emph{insulator} might exhibit \emph{conductive} behavior without charge excitations below the insulating gap presents a fascinating paradox. This research aims to postulate a potential resolution of this paradox, offering new insights into the behavior of the optical conductivity in SMG insulators and expanding our understanding of quantum many-body phenomena.

\emph{Pole-Zero Duality.}~---
Let us first revisit the Golterman-Shamir construction \cite{Golterman2024P2311.12790}, and reproduce their results through the broader lens of pole-zero duality in the fermion Green's function.

The analysis starts with the fermion two-point correlation function \(G(k) = -\langle\psi_k \psi_k^\dagger\rangle\) in the energy-momentum space. At this point, Golterman and Shamir introduced a pivotal \emph{assumption} that the fermion system can be approximated by a \emph{free} effective action,
\eq{S[\psi]=-\sum_{k}\psi_{k}^\dagger G(k)^{-1}\psi_{k}=-\int\dd^d x\;\psi^\dagger G(\ii\partial)^{-1}\psi,\label{eq:S}}
such that the Green's function will be consistently reproduced by the fermion path integral:
\eq{G(k)=\frac{1}{Z} \int \mathcal{D}[\psi] \, (-\psi_k \psi_k^\dagger) e^{-S[\psi]},} with \(Z = \int \mathcal{D}[\psi] e^{-S[\psi]} \) being the partition function. It is crucial to acknowledge that this approach to reconstructing the effective action from the two-point correlation is only valid under the premise that the fermions behave as free or generalized free fields \cite{Greenberg1961G,Dutsch2003Gmath-ph/0209035,Liu2018d1808.00612,Nebabu2023B2306.16687,Zeng2023B2309.03178}. In such cases, higher-point correlations decompose into two-point correlations via Wick's theorem. Should the fermions deviate from generalized free field behavior, it becomes necessary to incorporate higher-order terms in the effective action to model higher-point correlation functions.

If we accept the effective action $S[\psi]$ in \eqnref{eq:S}, we can proceed to gauge the $\U(1)$ symmetry of the fermion field $\psi$, under which \(\psi \to \e^{\ii \theta}\psi\). By introducing the $\U(1)$ gauge field \(A\) through minimal coupling, the effective action becomes:
\eq{S[\psi, A] = -\int\dd^dx\;\psi^\dagger G(\ii\partial-A)^{-1}\psi.\label{eq: SpsiA}}
Integrating out the fermion field $\psi$, the fermion path integral $\e^{-S[A]}=\int\scD[\psi]\e^{-S[\psi, A]}$ defines an effective action $S[A]$ for the gauge field $A$: 
\eqs{S[A]&=\Tr \log G(\ii\partial-A)\\
&=\sum_{n=0}^{\infty}\frac{1}{n!}\Pi_{n}^{\mu_1\mu_2\cdots\mu_n}A_{\mu_1}A_{\mu_2}\cdots A_{\mu_n},\label{eq: SA}}
where $\Pi_{n}^{\mu_1\mu_2\cdots\mu_n}:=\delta_{A_{\mu_1}}\delta_{A_{\mu_2}}\cdots \delta_{A_{\mu_n}}S[A]$ corresponds to the $n$th order current correlation, or loosely denoted as $\Pi_n=\delta_{A}^n S[A]$. These correlations $\Pi_n$ encode 
 the responses of the fermion system to the external electromagnetic field at different orders. The goal is to understand their behaviors across the SMG transition. 

The SMG transition refers to the fermion gap-opening transition driven by the fermion interaction \cite{Wang2204.14271}. On the weakly interacting side, the fermion system is metallic, characterized by gapless single-particle excitations at low energy, manifested as poles in the fermion Green’s function. In contrast, on the strongly interacting side, the system transitions into the SMG insulating phase, where the original poles in the Green's function are replaced by zeros. For instance, in the case of the relativistic fermions discussed in \refcite{Golterman2024P2311.12790}, the propagator poles and zeros are respectively modeled by $G_\text{Dirac}$ and $G_\text{SMG}$ as
\eq{G_{\text{Dirac}}(k) = \frac{1}{\gamma^0\gamma^\mu k_\mu}, \quad G_{\text{SMG}}(k) = -\frac{\gamma^0\gamma^\mu k_\mu}{m^2}.\label{eq: G Dirac SMG}}
The concept of pole-zero duality \cite{You1403.4938,Kaplan2021I2112.06954,Kaplan2022G2205.05707,Lu2023G2307.12223} offers a compelling framework for relating the low-energy behaviors of Green's functions across the SMG transition. This duality is articulated through a transformation of the fermion Green's function,
\eq{G(k) \to \tilde{G}(k)\propto G(-k)^{-1},\label{eq: dual}}
under which poles and zeros replace each other. For example, $G_\text{Dirac}$ and $G_\text{SMG}$ in \eqnref{eq: G Dirac SMG} are related by the pole-zero duality.

Under the pole-zero duality defined in \eqnref{eq: dual}, the effective gauge action $S[A]=\Tr\log G(\ii\partial-A)$ in \eqnref{eq: SA} transforms as
\eq{S[A]\to \Tr\log G(-\ii\partial+A)^{-1}=-S[-A],}
then the $n$th order current correlation $\Pi_n=\delta_{A}^nS[A]$ transforms as
\eq{\Pi_n\to -\delta_{A}^nS[-A]=-(-1)^n\Pi_n.}
This implies that if the fermion Green’s functions across the SMG transition are related by pole-zero duality, as exemplified by \( G_{\text{Dirac}} \) and \( G_{\text{SMG}} \) in \eqnref{eq: G Dirac SMG}, then their corresponding electromagnetic response functions will also be related by (for $n=2,3$)
\eq{\Pi_{2,\text{SMG}}=-\Pi_{2,\text{Dirac}},\quad \Pi_{3,\text{SMG}}=\Pi_{3,\text{Dirac}},} 
which reproduce the main conclusions in \refcite{Golterman2024P2311.12790} that, compared to free Dirac fermions, the vacuum polarization $\Pi_2$ changes sign in the SMG insulator while the triangle diagram $\Pi_3$ remains the same.

These results are remarkably general and do not depend on the specific form of the Green's function $G(k)$. Provided we accept the effective action $S[\psi, A]$ in \eqnref{eq: SpsiA} as our starting point, the conclusions outlined above are inevitable under the principle of pole-zero duality.

The implications of these results are significant. Since vacuum polarization is connected to optical conductivity by \(\Re \sigma(\omega,\vect{k}) = -\frac{1}{\omega}\Im \Pi_2(\omega,\vect{k})\) \cite{mahan2013many}, the relationship \(\Pi_{2,\text{SMG}} = -\Pi_{2,\text{Dirac}}\) would imply that \(\sigma_\text{SMG} = -\sigma_\text{Dirac}\), indicating that the gapped SMG insulator would exhibit a conductivity that is finite and opposite to that of the gapless Dirac semimetal. However, we should not anticipate finite conductivity in an insulator below the charge excitation gap. Additionally, the notion of negative conductivity raises concerns about the potential violation of the fluctuation-dissipation theorem and the loss of unitarity in the theory.

\emph{Unbounded Current Operator}~---
Given the perplexing behavior of conductivity, we are motivated to examine the foundational assumptions of the Golterman-Shamir construction. Specifically, it is presumed that the effective action $S[\psi,A]$ described in \eqnref{eq: SpsiA} models the physics of the SMG insulator in a background electromagnetic field. Starting from this premise, the current operator in the system should be given by
\eq{J^\mu=-\delta_{A_\mu}S[\psi,A\to 0]=\sum_{k}\psi_k^\dagger \partial_{k_\mu} G(k)^{-1}\psi_k.\label{eq: j}}
For Dirac fermions described by $G_\text{Dirac}(k)$ in \eqnref{eq: G Dirac SMG}, \eqnref{eq: j} gives the current operator in the conventional form $J^\mu=\sum_{k}\bar{\psi}_k\gamma^\mu\psi_k$ (where $\bar{\psi}:=\psi^\dagger \gamma^0$), which has a bounded spectrum. However, for the SMG insulator, if we naively substitute the Green's function $G_\text{SMG}(k)$ from \eqnref{eq: G Dirac SMG} into \eqnref{eq: j},
\eq{J^\mu=\sum_{k}\frac{m^2}{k^4}(2k_\mu k_\nu-k^2\delta_{\mu\nu})\bar{\psi}_k\gamma^\nu\psi_{k},}
we find that the resulting current operator $J^\mu$ would diverge as \( k^2:=k_\mu k^\mu \rightarrow 0 \). This divergence is a direct consequence of the Green's function zeros along the light cone (\(k^2= 0\)) below the SMG insulating gap. Such a current operator has an unbounded spectrum, which is unphysical because this would imply the existence of quantum states in which the velocity of charge movement could potentially exceed the speed of light.

Starting from such an unbounded current operator to define the current-current correlation \(\Pi_2^{\mu\nu} = -\langle J^\mu J^\nu\rangle\) could potentially lead to unphysical results. This perspective makes the unusual behavior of \(\Pi_2\) less surprising. It implies that \(S[\psi, A]\) in \eqnref{eq: SpsiA} might not be a complete theory for describing the SMG insulator. Given that the SMG insulator is intrinsically a strongly interacting system, it is reasonable to suspect that the action should include various higher-order terms to cancle the divergence of the current operator, thereby ensuring a well-defined bounded current operator.

\emph{Lattice Modeling}~---
Having recognized the critical issue with the unbounded current operator, our goal is to move beyond the effective action \( S[\psi, A] \) and explore the electromagnetic response of the SMG insulator from a more fundamental perspective.

To address this challenge, we turn to a concrete lattice model for the SMG insulator. Consider a system comprising four flavors of fermions \(c_{ia}\) (where \(a = 1, 2, 3, 4\)) on each lattice site \(i\), governed by the following Hamiltonian
\eq{
H = -\sum_{ij} t_{ij} \e^{\ii A_{ij}} c_{ia}^\dagger c_{ja} - g \sum_{i} c_{i1}^\dagger c_{i2}^\dagger c_{i3} c_{i4} + \text{h.c.},\label{eq: H}}
where the repeated flavor index $a$ in $c_{ia}^\dagger c_{ja}$ will be implicitly summed over, and ``$\text{h.c.}$'' represents the Hermitian conjugate terms. In this model, a background \(\U(1)\) gauge connection \(A_{ij}\) is introduced between every pair of sites to gauge the global $\U(1)$ symmetry of the fermions (acting as \(c_{ia} \to \e^{\ii\theta} c_{ia}\)). 

We further specify that the fermions in the system are \emph{half-filled}, a crucial condition for cancelling the Fermi-surface anomaly \cite{LuttingerRP1960,Paramekanticond-mat/0406619,Haldanecond-mat/0505529,Basar1307.2234,Watanabe1505.04193,Cheng1511.02263,Lu1705.09298,Cho1705.03892,Bultinck1808.00324,Song1909.08637,Else2007.07896,Wen2101.08772,Else2010.10523,Ma2110.09492,Wang2110.10692,Darius-Shi2204.07585,Cheng2211.12543} and enabling an SMG insulating state \cite{Lu2210.16304,Lu2302.12731}. To enforce the half-filling condition without fine-tuning the chemical potential, a simple approach is to first assume a bipartite lattice structure (e.g., a square or honeycomb lattice that can be partitioned into $A$ and $B$ sublattices), and then impose an anti-unitary sublattice particle-hole symmetry \( \dsZ_2^\scS \) (also known as the chiral symmetry \cite{Ryu0912.2157,Chiu1505.03535}), under which \( c_{ia} \to \scK (-)^i c_{ia}^\dagger \), with an alternating sign \( (-)^i = \pm1 \) for the site $i$ in $A$ and $B$ sublattices respectively. Here \( \scK \) represents the complex conjugation operator that $\scK^2=1$ and $\scK \ii \scK=-\ii$.

Let us first turn off the background gauge field by setting $A_{ij}= 0$. In the free fermion limit (\(g = 0\)), the Hamiltonian \(H\) describes a fermion hopping model on a bipartite lattice with a chiral symmetry $\dsZ_2^\scS$. In the momentum space, the $\dsZ_2^\scS$ symmetry transforms the fermions as $c_{\vect{k}a}\to \scK \sigma^3 c_{\vect{k}a}^\dagger$, enforcing the Hamiltonian to take the form \(H = \sum_{\vect{k}} c_{\vect{k}a}^\dagger \xi_{\vect{k}}\sigma^1 c_{\vect{k}a}\), where $\sigma^{\alpha}$ ($\alpha=0,1,2,3$) denote the Pauli matrices acting within the sublattice Hilbert space. The specific details of the band dispersion \(\xi_{\vect{k}}\) are not crucial to our discussion. Without fine-tuning the band structure, $\xi_\vect{k}$ typically exhibits a Fermi surface, rendering the fermion system as a gapless Fermi liquid in general.

Conversely, in the strong interaction limit (\(g \to \infty\)), the hopping term $t_{ij}$ can be omitted relative to $g$, and the model is decoupled to individual sites, permitting an independent solution for each site. In this limit, the exact many-body ground state of $H$ is a product state: 
\eq{
\ket{\Psi_\text{SMG}} = \prod_{i} \frac{1}{\sqrt{2}} (c_{i1}^\dagger c_{i2}^\dagger - c_{i3}^\dagger c_{i4}^\dagger) \ket{0},\label{eq: Psi SMG}}
where \(\ket{0}\) represents the vacuum state of the fermions. This solution arises because the four-fermion interaction $g$ directly hybridizes the two-fermion states \(c_{i1}^\dagger c_{i2}^\dagger\ket{0}\) and \(c_{i3}^\dagger c_{i4}^\dagger\ket{0}\) on each site, thereby lowering the energy of the particular superposition state of them in \eqnref{eq: Psi SMG}. The resulting product state \(\ket{\Psi_\text{SMG}}\)  maintains the full symmetry of the Hamiltonian \(H\) and exhibits a gap of order \(g\) to all excitations, therefore realizing an SMG insulator (in its ideal limit). It is noteworthy that the fermions are automatically half-filled on every site, which is precisely why we emphasize the half-filling condition from the outset. Otherwise, we would have to violate the fermion number conservation when driving the system into the SMG insulating state as we increase the interaction strength $g$. 

\emph{Ideal SMG Limit.}~---
As long as at half-filling, regardless of the band structure of $\xi_{\vect{k}}$, strong enough interaction $g$ in this model will always drive the fermion system into the SMG insulating phase. Understanding how the SMG transition happens, as a metal-insulator transition, is a fascinating yet challenging problem. However, this inquiry is beyond the scope of our current analysis. Instead, our focus will be on the strongly interacting regime where \( g \gg t_{ij} \), and we aim to study the current correlation in the SMG insulating phase. 

First, to properly define the current operator in the lattice model \eqnref{eq: H}, we start by introducing the background gauge field \(A_{ij}\), differentiating \(H\) with respect to \(A_{ij}\) and subsequently taking the limit of \(A_{ij} \to 0\),
\eq{
J_{ij} = \frac{\delta H}{\delta A_{ij}} = -(\ii t_{ij} c_{ia}^\dagger c_{ja} + \text{h.c.}).\label{eq: Jij}}
This current operator $J_{ij}$, well-defined on the lattice, has a bounded spectrum and does not suffer from the previous problem of unbounded current in the effective action approach. We can observe that the on-site interaction \( g \) has no influence on the definition of the current operator in \eqnref{eq: Jij}, as the interaction term was not modified by the \(\U(1)\) background field $A_{ij}$ in the Hamiltonian \eqnref{eq: H} to begin with. The lattice current operator \(J_{ij} \) is entirely determined by the hopping term. Therefore, in momentum space, the current operator \( \vect{J}= \sum_\vect{k}c_{\vect{k}a}^\dagger \partial_\vect{k}\xi_\vect{k}\sigma^1 c_{\vect{k}a}\) can also be expressed as solely dependent on the band dispersion $\xi_\vect{k}$, the same as in the free fermion limit.

Then, we can compute the current-current correlation on the lattice as
$\Pi(t)=-\ii\langle [J_{ij}(t),J_{kl}(0)]\rangle\Theta(t)$, where $J_{ij}(t)=\e^{-\ii H t} J_{ij} \e^{\ii H t}$. Evaluating the operator expectation values on the ideal ground state $\ket{\Psi_\text{SMG}}$ in  \eqnref{eq: Psi SMG} of SMG insulator in the $g\gg t_{ij}$ limit, we find $\Pi(t)=4|t_{ij}|^2\sin(2 g t)\Theta(t)\delta_{il}\delta_{jk}$. Fourier transform to the frequency domain and averaging over all sites, the optical conductivity $\Re\sigma(\omega)=-\frac{1}{\omega}\Im\Pi(\omega)$ reads
\eq{\Re\sigma(\omega)=\frac{2\pi |t|^2}{2g}\big(\delta(\omega-2g)+\delta(\omega+2g)\big),}
where $|t|^2:=\sum_{j}|t_{ij}|^2$ characterizes the overall hopping strength. This result illustrates the expected reasonable behavior of the conductivity in an ideal SMG insulator: $\Re\sigma(\omega)$ should vanish at low frequencies \( |\omega| < 2g \) within the charge gap \( 2g \). The presence of sharp peaks at $\omega=\pm 2g$ is attributed to the fact that all excitations exhibit flat dispersion in the ideal SMG state, as the system decouples among independent sites. Deviating from this ideal limit, for a finite \( t_{ij}/g \), we should anticipate the peaks to broaden into continua above the charge gap. 

\emph{Beyond Ideal Limit.}~---
To elucidate the general behavior of optical conductivity in the SMG insulator beyond the strong interaction limit, we engage a perturbative expansion in \(t_{ij}/g\) around the ideal state \(\ket{\Psi_\text{SMG}}\) and calculate the correlation functions in the perturbed state. The perturbation theory has been thoroughly analyzed in \refcite{Lu2023G2307.12223}. The results indicate that the fermion Green’s function in the SMG insulator of a $\dsZ_2^\scS$-symmetric two-band system can be approximated by:
\eq{G_\text{SMG}(k) = \frac{\omega \sigma^0 + \xi_\vect{k}\sigma^1}{\omega^2-\xi_\vect{k}^2-\Delta^2},\label{eq: G SMG}}
where \(\Delta\sim g\) represents the single-particle gap (i.e., the fermion mass) renormalized from its bare value $g$ set by the interaction strength. It is worth emphasizing that the gap $\Delta$ here is not a symmetry-breaking order parameter and has no mean-field interpretation. Notably, $G_\text{SMG}(k)$ exhibits zeros along $\omega=\pm\xi_{\vect{k}}$, a distinguishing feature of SMG insulators. By contrast, in systems where the gap opens due to spontaneous symmetry breaking (SSB), the fermion Green’s function would take the following form:
\eq{
G_\text{SSB}(k) = \frac{\omega \sigma^0 + \xi_\vect{k}\sigma^1+\Delta \sigma^3}{\omega^2-\xi_\vect{k}^2-\Delta^2},\label{eq: G SSB}}
where the extra term $\Delta\sigma^3$ on the numerator breaks the $\dsZ_2^\scS$ symmetry, which is particularly absent in $G_\text{SMG}$. 

In any scenario, the charge \(\Pi^{00}\) and current \(\Pi^{ij}\) correlation functions can be characterized by the following equations:
\eqs{
\Pi^{00}(q) &= \Tr(G(k+q) G(k)), \\
\Pi^{ij}(q) &= \Tr(v_i G(k+q) v_j G(k)).\label{eq: Pi}}
In these expressions, the charge vertex operator is always \(\sigma^0\), and the current vertex operator is \(\vect{v} = \partial_{\vect{k}}\xi_{\vect{k}}\sigma^1\). These vertex operators align with the lattice current operators derived in \eqnref{eq: Jij}, ensuring a bounded operator spectrum.

\begin{figure}
\begin{center}
\includegraphics[width=0.95\linewidth]{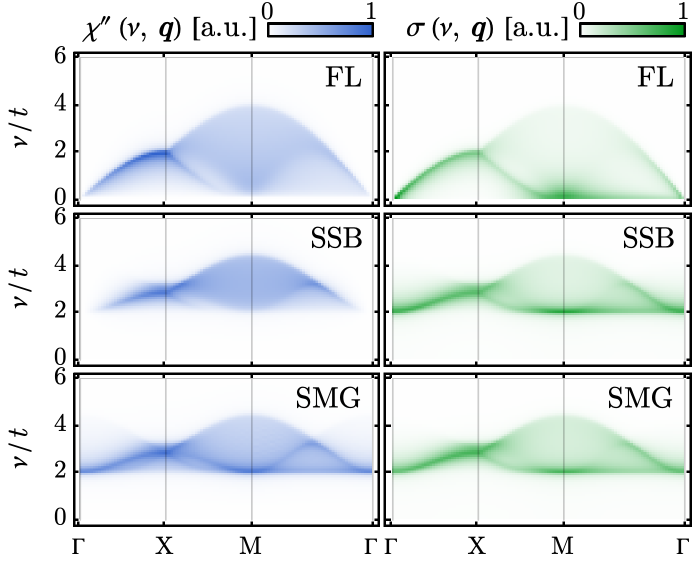}
\end{center}
\caption{The dynamic charge susceptibility $\chi''(\nu,\vect{q})$ (imaginary part) and optical conductivity $\sigma(\nu,\vect{q})$ as a function of frequency $\nu$ along a cut through  $\Gamma(0,0)$, X$(\pi,0)$, M$(\pi,\pi)$ points in the momentum space, for 2D square lattice fermions in the Fermi liquid (FL), spontaneous symmetry breaking (SSB), and symmetric mass generation (SMG) phases respectively.}
\label{fig: response}
\end{figure}

For example, consider a square lattice fermion model characterized by the dispersion relation $\xi_\vect{k}=-2t(\cos k_x+\cos k_y)$ given the hopping parameter $t$ (set as the energy unit). We can numerically evaluate the spectral weight of the dynamic charge susceptibility $\chi''(\nu,\vect{q})=-2\Im\Pi^{00}(\nu+\ii 0_+,\vect{q})$ and the optical conductivity $\sigma(\nu,\vect{q})=-\frac{1}{\nu}\Im\Pi^{ii}(\nu+\ii 0_+,\vect{q})$ as outlined in \eqnref{eq: Pi}. Plugging in the fermion Green's function in different phases: $G_\text{FL}(k)=(\omega\sigma^0-\xi_\vect{k}\sigma^1)^{-1}$, $G_\text{SSB}(k)$ in \eqnref{eq: G SSB}, and $G_\text{SMG}(k)$ in \eqnref{eq: G SMG} (assuming $\Delta=t$ in the SSB and SMG cases), their resulting electromagnetic response functions are compared in \figref{fig: response}. Much like that in the SSB insulator, the SMG insulator's electromagnetic response is gapped, displaying only subtle differences in detail. Contrarily, it does not resemble the gapless response typical of a Fermi liquid.

This analysis can be extended to SMG in non-chiral Dirac/Weyl fermions, where on-site local gapping interactions generally exist, ensuring that the current operator remains unaffected by interactions. This is a crucial element for our argument concerning the lattice regularization of the current operator. However, for chiral fermions, such as in the 3-4-5-0 model \cite{Wang1807.05998,Wang1809.11171,Zeng2202.12355,van-Beest2023M2306.07318}, where the SMG interaction is not on-site, regularizing the current operator is an open problem for future research. Thus, while the paradox regarding how the SMG insulator can exhibit finite optical conductivity despite its gap is effectively resolved for non-chiral fermions, further work is needed for chiral systems. This resolution hinges on a nuanced understanding of the current operator in the SMG phase using lattice regularization. These insights reinforce the SMG state's insulating nature while clarifying its distinctive low-energy electromagnetic properties, laying a theoretical foundation for future experimental exploration of featureless Mott insulators.

\begin{acknowledgments}
We acknowledge the discussions with Simon Catterall, Cenke Xu, Srimoyee Sen, David Tong, Yigal Shamir, Maarten Golterman, and Lei Su. We thank Lei Su for sharing their related upcoming work \cite{SuG} with us. MZ, DCL, and YZY are supported by a startup fund by UCSD and the National Science Foundation (NSF) Grant No. DMR-2238360. FX is supported by the Zhenggang Scholarship Program at Nanjing University. This research was supported in part by grant NSF PHY-2309135 to the Kavli Institute for Theoretical Physics (KITP), through the KITP Program ``Correlated Gapless Quantum Matter'' (2024). We acknowledge the OpenAI GPT4 for providing language and editing suggestions throughout the process of writing this paper.
\end{acknowledgments}

\bibliographystyle{apsrev4-2}
\bibliography{ref}

\end{document}